\documentstyle[twoside,fleqn,espcrc2]{article}
\newcommand{\be}{\begin{equation}}
\newcommand{\ee}{\end{equation}}
\newcommand{\bea}{\begin{eqnarray}}
\newcommand{\eea}{\end{eqnarray}}
\newcommand{\no}{\noindent}

\newcommand{\Tr}{{\rm Tr}}

\def\NPB{{Nucl.\ Phys.} B}
\def\PLB{{Phys.\ Lett.} B}

\newcommand{\AmS}{{\protect\the\textfont2
  A\kern-.1667em\lower.5ex\hbox{M}\kern-.125emS}}

\title{
   Cooling, Physical Scales and the Vacuum Structure of  Y-M Theories.
}

\author{
Margarita Garc\'{\i}a P\'erez\address{Departamento de F\'{\i}sica 
Te\'orica, Univ. Aut\'onoma de Madrid, Canto Blanco, 28049 Madrid, Spain},
Owe Philipsen\address{Theory Division, CERN, CH-1211 Geneva 23, Switzerland}
and 
Ion-Olimpiu Stamatescu\address{FESt, Schmeilweg 5, D-69118 Heidelberg
 and ITP, Philosophenweg 16,
D-69120 Heidelberg, Germany}\thanks{Partial support from the European Network ``Finite 
Temperature Phase Transitions in Particle Physics".} }\date{}

\begin{document}

\begin{abstract}
We present a cooling method controlled by a physical {\it cooling radius}
that defines a scale below which fluctuations are smoothed out while leaving
physics unchanged at all larger scales. This method can be generally used 
 as a gauge invariant low pass filter to extract the physics from
noisy MC configurations. Here
we apply this method to study topological properties of lattice
gauge theories where it allows to retain 
instanton--anti-instanton pairs.
\end{abstract}

\maketitle



\no Since MC configurations are noisy at small distances, 
smoothing is needed to observe the physical structure. 
One approach is ``cooling" 
  \cite{tep0}, a local minimization of a given lattice action. 
Cooling works as a diffusion process, smoothing out
 increasingly large regions
 \cite{dig89}. Thereby 
the physical spectrum is affected, in particular the string tension 
 drops
rapidly with cooling. Topological excitations
 may be distorted by bad scaling properties of the
action and by instanton - anti-instanton (I-A) annihilation
(pairs are not minima of the action). We thus need:\par
\no 1) to use an action with
 practically scale invariant instanton solutions and a dislocation threshold
to eliminate UV noise (dislocations), and\par
\no 2) to control cooling by means of a physical scale such that no monitoring or 
engineering which may
introduce uncertainties is
necessary.\par 
\no {\it Restricted Improved Cooling} (RIC) fulfills these requirements \cite{ric}.
RIC preserves physics at scales above a 
{\it cooling radius} $r$ which can be fixed unequivocally
beforehand,
while smoothing out the structure below $r$.
In particular the string tension is preserved beyond $r$,
instantons are stable, dislocations are eliminated, and I-A
 pairs are retained above a threshold 
defined by $ r $.\par\smallskip

\no {\bf Properties of RIC.}$\ $
RIC uses the action of the {\it Improved Cooling} algorithm (IC)
with 5 planar, fundamental Wilson loops \cite{ic}.
This action is 
 correct to order ${\cal O}(a^6)$ 
and has a {\it dislocation threshold} 
$\rho_0 \sim 2.3a$, below which short range topological structure 
is smoothed out (note that $\rho_0 \rightarrow 0$ in approaching continuum).
Above  $\rho_0$, 
instantons are stable to
any degree of cooling (however, I-A pairs annihilate).  The corresponding
{\it improved charge density} using the same combination of loops
leads to an integer charge already after a few cooling
sweeps and  stable thereafter.

Here we shall restrict ourselves to $SU(2)$, for the general case 
see \cite{ric}. Recall that the cooling  
algorithm is derived from the equations of motion
\be
\label{eq}
U_\mu(x) W_\mu(x)^\dagger - W_\mu(x) U_\mu(x)^\dagger = 0  ,
\ee
\no where $W$ is the sum of staples connected to the link $U_\mu(x)$ 
in the action. It amounts to substitute
\be
U \rightarrow U' = V = \frac{W}{||W||} \ , \ ||W||^2=\frac{1}{2} \Tr(W W^\dagger).
\label{e.sta}
\ee
\no We define RIC by the  constraint\footnote{We thank 
F. Niedermayer for this suggestion.}
that only those links be updated, which violate the equation of motion by
more than some
chosen threshold:
\be
U \rightarrow V \ \  {\rm iff} \ \  
\Delta_\mu(x)^2= \frac{1}{a^6} {\rm Tr}(1-U V^{\dagger})\geq\delta^2 .
\label{cond}
\ee
\no We have
$\Delta_\mu^2 (x) \propto 
-\Tr((D_\nu F_{\nu \mu}(x))^2)$ 
 in  continuum limit \cite{map}. 
 Thus 
$\delta$ controls the energy 
of the fluctuations around classical solutions 
and acts as a filter for short wavelengths. 
Since it uses the same action RIC has the same scaling properties as IC. 
However, since RIC  does not update links  already close to a solution,
it changes fewer links after
every iteration until  the algorithm saturates.
(\ref{cond}) defines a constrained minimization and 
the smoothing is 
 homogeneous over the lattice.

The next step is to relate the parameter $\delta$ which defines the 
cooling to a physical scale. For Yang Mills theory the latter should
 involve the string tension $\sigma$. We calculate the ``effective mass"
 $M(t)$ from  
correlation functions of spatial
Polyakov loops separated by $t$ steps in time. Asymptotically 
$M(t) \simeq N_s\sigma$ up to finite size corrections.

For the calculations we generate configurations using the
 Wilson action on the lattices:\par
 $12^3\times 36$, p.b.c., $\beta= 2.4$ ($a=0.12$ fm): 
800 configurations with 100 sweeps separation; \par
 $24^4$, twist in time, $\beta=2.6$ ($a= 0.06$ fm): 350 configurations 
with 200  sweeps separation.\par
\no We do 20000 thermalization sweeps. We use $\delta=2.89,~4.92,~11.57,~23.15$ 
and $46.30$ fm$^{-3}$.
 
\begin{figure}[htb]
\vspace{2.5cm}
\includegraphics{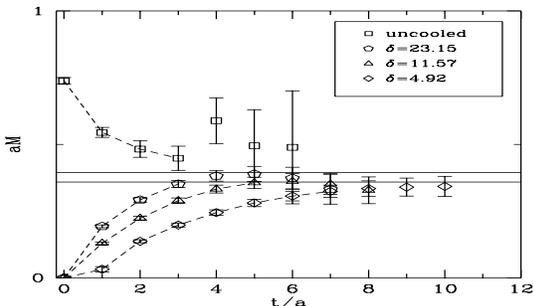} 
\caption[]{\label{f.mef}
{\it $M(t)$ for the $a=0.12$ fm lattice. The horizontal band represents 
standard results for $\sigma$.
}}
\vspace{-0.8cm}
\end{figure}

\begin{figure}[htb]
\vspace{3cm}
\includegraphics{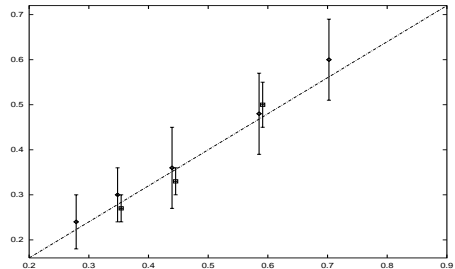}
\caption[]{\label{f.rde}
{\it The smoothing scale of RIC: $r(\delta)$ vs $\delta^{-1/3}$ 
for the $a=0.12$ fm
(diamonds) and $a=0.06$ fm (squares) lattices (the middle 
points are slightly
displaced around the true positions for readability). Also
shown is the
fit (\ref{e.cscale}).
}}
\vspace{-0.8cm}
\end{figure}

For illustration we present in Fig. \ref{f.mef} $M(t)$ for the $24^4$ lattice.
Defining  $r(\delta)$
as the distance $t$ at which $M(t)$ on $\delta$-cooled configurations
starts to agree with the uncooled value (obtained by fuzzing and fitted 
to a smooth function of $t$) we arrive at  
the results in Fig. \ref{f.rde}, compatible with
\be
r(\delta) \simeq 0.8 \ \delta^{-1/3} ,  
\label{e.cscale}
\ee
\no showing the correct scaling behaviour, as expected since  $\delta$ has
a continuum limit.

The behaviour of instantons is similar under RIC and IC whereby the former
has a 
somewhat smaller, $\delta$-dependent dislocation threshold 
$\rho_0(\delta)$. I-A pairs, however, do not annihilate under RIC the way they 
do for the other cooling methods (including IC): 
since the distortion of the 
partners in a pair depends on the overlap,
RIC preserves pairs below some overlap-threshold. More precisely, 
it turns out that there is a well defined 
relation between $\Delta$
and $S_{\rm int}^{\rm IA}=16\pi^2-S^{IA}$,
 therefore RIC   
stabilizes I-A pairs with $S_{\rm int}$ below  a 
threshold which is a function of $\delta$, 
hence of $ r$.\par\smallskip

\no {\bf SU(2) topological properties by RIC.} The total, improved
topological charge stabilizes to integer values after only few IC 
or RIC sweeps. Therefore the topological susceptibility appears very 
stable if $\delta$ is small enough to eliminate dislocations,  shows correct scaling
and agrees with the Witten-Veneziano relation.

To describe the details of the
instanton ensemble one needs to recognize the particular
I's and A's. Since the objects can be very distorted,
our description relies on two assumptions:
a) instantons should appear as local self-dual peaks in the action and charge
density, and b) only pairs with $S_{\rm int}^{\rm IA}$ 
considerably smaller than $16 \pi^2$ 
should be considered as such. 
We approximate the action and charge
density by a superposition of I's and 
A's parameterized through the 
BPST formula for an instanton of size $\rho$. 
We  measure the adequacy of this  
 ansatz through the fraction of the total action (charge) reproduced
in this way. We represent the data 
in physical units using the corresponding $a$ and the 
$r(\delta)$ obtained from the string tension analysis 
without  any further fit
or tuning.

\begin{figure}[htb]
\vspace{4.3cm}
\includegraphics{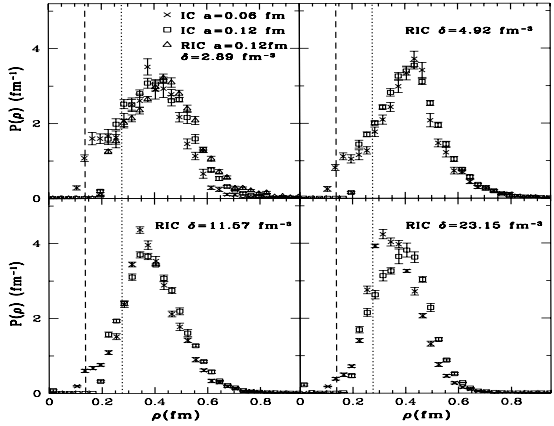}
\caption{{\it Normalized size 
distributions on the $a=0.12$ fm lattice (squares, triangles)
and $a=0.06$ fm lattice (crosses).
Vertical dotted (dashed) lines indicate the  IC dislocation
threshold $\rho_0=2.3a$ for $a=0.12$ ($0.06$) fm.
}}
\label{f.size}
\vspace{-0.8cm}
\end{figure}

\begin{figure}[htb]
\vspace{3cm}
\includegraphics{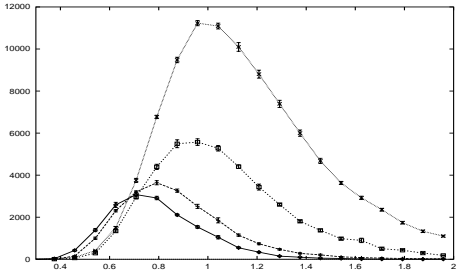}
\caption{{\it Non-normalized overlap 
distributions for  $a=0.06$ fm and $\delta=2.89$ (circles), $4.92$ (bars),
$11.57$ (squares) and $23.15$ (crosses) fm$^{-3}$.
}}
\label{f.over}
\vspace{-0.8cm}
\end{figure}

The main results of our analysis are:\par
\no 1. The density of instantons increases drastically with 
decreasing smoothing scale $r$.\par
\no 2. Size distributions depend only weakly on $r$.\par
\no 3. The typical I-A distance $d_{IA}$ (from an I to the nearest A) 
seems given by the smoothing scale.\par
\no 4. The overlap $(\rho_I+\rho_A)/2d_{IA}$ 
increases strongly with decreasing $r$ and 
the fit quality deteriorates.\par
\begin{figure}[htb]
\vspace{3cm}
\includegraphics{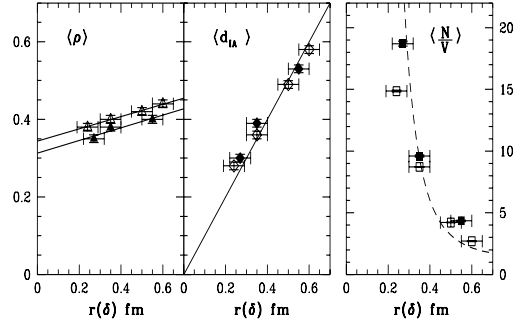}
\caption[]{\label{f.extr}
{\it Dependence of the average size (triangles),
the average I - A  distance (diamonds)
and the instanton density (squares)
on the cooling radius $r$. 
The dashed curve is 
$\langle N/V\rangle= \langle N/V\rangle_{\rm asymptotic}
+ C/r^4$.
Open (filled) symbols correspond to the $a=0.12$ $(0.06)$ fm lattice.
}}
\vspace{-0.6cm}
\end{figure}

\no All these results are largely invariant under rescaling of the
lattice spacing $a$. They are summarized in the Figures \ref{f.size}, 
\ref{f.over} and \ref{f.extr}.

It appears therefore that there are topological 
features of the vacuum structure, like susceptibility and charge distributions 
and to a fair degree 
also size distributions, which 
are well defined, show correct scaling and are rather independent on the 
smoothing scale once the noise has been  damped sufficiently. However,
I(A)-density, overlap and I-A distance distributions are not of this kind:
although these quantities seem to scale correctly with the cut off, their 
behaviour with $r$ suggests that looking at smaller and smaller scales a continuous 
spectrum of fluctuations emerges and a
description in terms of I's, A's and I-A pairs 
becomes less and less meaningful. This may explain both the
agreements and the disagreements between various analyses -- 
see, e.g. \cite{oth}.

Generally RIC proves itself a good instrument for 
probing different scales in a controlled way
and extract physics from noisy configurations.

\end{document}